\documentclass[]{emulateapj}

\usepackage{amsmath}
\usepackage{dcolumn,multirow,ccaption}
\usepackage{graphicx}
\usepackage{color}
\usepackage{ccaption}
\usepackage{lscape}
\usepackage[T1]{fontenc}
\usepackage{textcomp}
\usepackage{ulem}

\shorttitle{
Star Formation On Sub-kpc Scale Triggered By Non-linear Processes In Nearby Spiral Galaxies
}
\shortauthors{Momose et al.}

\begin{document}

\title{Star Formation On Sub-kpc Scale Triggered By Non-linear Processes In Nearby Spiral Galaxies}

\author{Rieko Momose\altaffilmark{1,2,3}, Jin Koda\altaffilmark{4}, Robert C. Kennicutt, Jr\altaffilmark{5}, Fumi Egusa\altaffilmark{6}, Daniela Calzetti\altaffilmark{7}, Guilin Liu\altaffilmark{7,8}, Jennifer Donovan Meyer\altaffilmark{4,9}, Sachiko K. Okumura\altaffilmark{3,10}, Nick Z. Scoville\altaffilmark{11}, Tsuyoshi Sawada\altaffilmark{3,12}, Nario Kuno\altaffilmark{13}}

\altaffiltext{1}{Current address: Institute for Cosmic Ray Resarch, University of Tokyo, 5-1-5 Kashiwano-ha, Kashiwa city, Chiba 277-8582, Japan; momo@icrr.u-tokyo.ac.jp}
\altaffiltext{2}{Department of Astronomy, University of Tokyo, Hongo, Bunkyo-ku, Tokyo 113-0033, Japan; momo.s.rieko@nao.ac.jp}
\altaffiltext{3}{National Astronomical Observatory of Japan, Mitaka, Tokyo 181-8588, Japan}
\altaffiltext{4}{Department of Physics and Astronomy, Stony Brook University, Stony Brook, NY 11794-3800, USA}
\altaffiltext{5}{Institute of Astronomy, University of Cambridge, Cambridge CB3 0HA, UK}
\altaffiltext{6}{Institute of Space and Astronautical Science, Japan Aerospace Exploration Agency, Chuo-ku, Sagamihara, Kanagawa 252-5210, Japan}
\altaffiltext{7}{Astronomy Department, University of Massachusetts, Amherst, MA 01003-9305, USA}
\altaffiltext{8}{Current address: Center for Astrophysical Sciences, Johns Hopkins University, 3400 North Charles Street, Baltimore, MD 21218-2686, USA}
\altaffiltext{9}{Current address: National Radio Astronomy Observatory, 520 Edgemont Rd., Charlottesville, VA, USA}
\altaffiltext{10}{Current address: Department of Mathematical and Physical Sciences, Japan Women's University, 2-8-1 Mejirodai, Bunkyo-ku, Tokyo 112-8681, Japan}
\altaffiltext{11}{Astronomy Department, California Institute of Technology, MC 249-17, 1200 East California Boulevard, Pasadena, CA 91125, USA}
\altaffiltext{12}{Joint ALMA Observatory, Alonso de C$\'{o}$rdova 3107, Vitacura, Santiago 763-0355, Chile}
\altaffiltext{13}{Nobeyama Radio Observatory, National Astronomical Observatory of Japan, Nobeyama, Minamimaki, Minamisaku, Nagano, Japan 384-1305}

\begin{abstract}
We report a super-linear correlation for the star formation law based on new CO($J$=1-0) data from the CARMA and NOBEYAMA Nearby-galaxies (CANON) CO survey. The sample includes 10 nearby spiral galaxies, in which structures at sub-kpc scales are spatially resolved. Combined with the star formation rate surface density traced by H$\alpha$ and 24 $\mu$m images,  CO($J$=1-0) data provide a super-linear slope of $N$ = 1.3. The slope becomes even steeper ($N$ = 1.8) when the diffuse stellar and dust background emission is subtracted from the H$\alpha$ and 24 $\mu$m images. In contrast to the recent results with CO($J$=2-1) that found a constant star formation efficiency (SFE) in many spiral galaxies, these results suggest that the SFE is not independent of environment, but increases with molecular gas surface density. We suggest that the excitation of CO($J$=2-1) is likely enhanced in the regions with higher star formation and does not linearly trace the molecular gas mass. In addition, the diffuse emission contaminates the SFE measurement most in regions where star formation rate is law. These two effects can flatten the power law correlation and produce the apparent linear slope. The super linear slope from the CO($J$=1-0) analysis indicates that star formation is enhanced by non-linear processes in regions of high gas density, e.g., gravitational collapse and cloud-cloud collisions.
\end{abstract}


\section{introduction}
The relation between star formation rate surface density ($\Sigma_{\text{SFR}}$) and gas surface density ($\Sigma_{\text{gas}}$) is important for understanding of the star formation mechanism in galaxies. \citet{sch59} suggested a power law correlation between $\Sigma_{\text{SFR}}$ and $\Sigma_{\text{gas}}$, $\Sigma_{\text{SFR}} \propto \Sigma_{\text{gas}}^N$. \citet[hereafter  K98a, K98b]{ken98a, ken98b} expanded the early study substantially and found a super-linear correlation with $N$ = 1.4 by combining atomic (HI) and molecular (CO $J$=1-0) data, and calculating SFR using H$\alpha$ and infrared data. This correlation is often called the Kennicutt-Schmidt law (hereafter, the K-S law). Extensive reviews can be found in K98b and \citet{ken12}.

The index of the K-S law ($N$) has been a primary indicator of the mechanism of star formation. Recent studies of the K-S law investigated the index at sub-kpc scales, approaching the intrinsic scale of star formation, i.e., the sizes of giant molecular clouds (GMCs) or giant molecular associations (GMAs). \citet[][hereafter B08]{big08} derived $N$ $\sim$ 1 from the correlation between $\Sigma_{\text{SFR}}$ and the molecular gas surface density $\Sigma_{\text{H$_2$}}$ estimated from CO($J$=2-1) data for 7 nearby spiral galaxies. They suggested that a linear correlation is evident in regions of high gas surface densities where the gas is typically molecular ($\geq$ 10 M$_\odot$ pc$^{-2}$). Several studies using CO($J$=2-1) also showed a linear correlation (e.g. \citealp{ler08,sch11}), including one which combined single-dish CO($J$=2-1) data with interferometric CO($J$=1-0) data (\citealp{rah11}). These CO($J$=2-1) studies analyzed a substantial number of nearby galaxies, though it should be recognized that some studies based on CO($J$=1-0) data showed a super-linear (power-law) correlation, rather than a linear correlation (e.g. \citealp{won02, ken07,liu11}).

CO($J$=1-0) is a better calibrated tracer of the bulk molecular gas in spiral galaxies. While CO(2-1) emission is sensitive to density and temperature changes \citep{kod12}, the CO(1-0)-to-H$_2$ conversion factor is more stable since their changes compensate each other to some extent \citep{sco87} as empirically demonstrated \citep{bol08, bol13}. However there is no systematic study of the K-S law on sub-kpc scale using CO($J$=1-0) for a large number of galaxies. A recent study by \citet{rah12} used CO($J$=1-0) data from an interferometer (but no single-dish data). The flux measured by an interferometer is reliable only when an object of interest is very compact and isolated. In any extended structures and their surrounding areas (e.g., spiral arms, inter-arms), the measured fluxes are uncertain. Without access to single-dish data, one doesn't know how much flux is resolved out in the interferometry. Such a systematic bias hinders a study of the K-S law with interferometer-only data, as star-forming regions are located predominately along spiral arms (e.g. \citealp{sco01}).

We study the K-S law on sub-kpc scales using new CO($J$=1-0) data obtained by the Nobeyama 45m single-dish telescope (NRO45) and CARMA. In this letter, we discuss only the correlation of $\Sigma_{\text{SFR}}$ with $\Sigma_{\text{H$_2$}}$, but not with the total gas surface density $\Sigma_{\text{gas}}$. The gas phase is predominantly molecular in regions where $\Sigma_{\text{gas}}$ is above 10 M$_\odot$ pc$^{-2}$ (B08), and the correlation holds even when $\Sigma_{\text{H$_2$}}$ alone is analyzed (e.g. \citealp{ken07}, B08). 

In this paper, we find a non-linear index of the K-S law in our spatially resolved analyses (on sub-kpc scales) and discuss the star formation mechanisms. We describe the data and methods of this study in $\S$ 2 and present results in $\S$ 3. In $\S$ 4, we compare our new results with the previous studies and discuss star formation mechanisms.

\section{Data and Method of The Study}
\subsection{Molecular Gas Surface Density}
We use CO($J$=1-0) data from the CARMA and NOBEYAMA Nearby-galaxies (CANON) CO($J$=1-0) survey \citep[Koda et al. in preparation; see also ][]{don13}. This survey combines high-resolution interferometer data (CARMA) and total power single-dish data (NRO45), providing high-fidelity, high-resolution CO($J$=1-0) data for nearby spiral galaxies. The single-dish data are critical to the flux measurements. For instance, the flux in a CARMA-alone map varies systematically with galactic structures; in the case of M 51 (\citealp{kod11}), the recovered flux is as low as $\sim$ 20 $\%$ in inter-arm regions and as high as $\sim$ 100 $\%$ in spiral arms.

We analyze 10 galaxies from the CANON sample (NGC 3521, 3627, 4254, 4303, 4321, 4736, 4826, 5055, 5194 and 6946). We convert the CO($J$=1-0) integrated intensity to $\Sigma_{\text{H$_2$}}$ using the standard $X_{\text{CO}}$ = 2.0 $\times$ 10$^{20}$ cm$^{-2}$ (K km s$^{-1}$)$^{-1}$ (\citealp{sto96, dam01}), since all our samples are late-type spiral galaxies whose metallicities are nearly the value of solar metallicity (\citealp{mou10}).

\subsection{Star Formation Rate}
We combine H$\alpha$ and 24 $\mu$m images to estimate $\Sigma_{\text{SFR}}$ using Equation (7) in \citet{cal07}. Most H$\alpha$ and all 24 $\mu$m images are obtained from  the Spitzer Infrared Nearby Galaxies Survey (SINGS) archive (\citealp{ken03}). We also use some H$\alpha$ images (NGC 4303, 4736 and 4826) obtained via the NASA Extragalactic Database, NED (\citealp{kna04, dal09}).

We discuss the effect of local background (BG) emission on the SFR estimate and investigate the K-S law with and without BG subtraction. The origin of H$\alpha$ BG may be leaked photons from distant HII regions (a.k.a. diffuse ionized gas, ``DIG'', e.g. \citealp{ferg96}). The BG of 24 $\mu$m emission may be the radiation from small dust grains heated by old stars (e.g. \citealp{dra07}).
Such BG emission is not related to recent star formation events, and therefore, if present, should be removed in the SFR estimate.
The BG subtraction affects the index of the K-S law significantly. The BG subtraction is not quite straightforward, and various techniques have been suggested (e.g. \citealp{rah11,liu11}). In this study we adopt the technique introduced by \citet{liu11}, in which they divided HII regions and the diffuse BG emission using the publicly-available IDL routine $HIIphot$ (\citealp{thi00}), and used the output image as BG image which is generated by the interpolation from the BG pixels around HII regions and surface-fitting to estimate BG emission embedded in the HII regions. We also carry out smoothing of the BG image the same manner as \citet{liu11}. 

We find the BG fractions of H$\alpha$ and 24 $\mu$m to be 33-57 $\%$ and 39-66 $\%$ of the total luminosities, respectively. These are consistent with previous results [$\sim$30-50 $\%$ Lyman photon leakage from HII regions (\citealp{ferg96}); and the 24  m BG of 30-40 $\%$ in galactic centers and 20 $\%$ in disks (\citealp{ver09})]. \citet{dra07a} also found   88 $\%$ of infrared emission coming from regions of very low stellar radiation fields, presumably not associated with local star formation, while \citet{ler12} found a smaller fraction ($\sim$ 20 $\%$). There may be some room for debate on the actual BG fractions (e.g., dependences on local radiation field, the amount of underlying dust, dust grain compositions, etc). However, their effects on the index $N$ is simple and systematic. This will be discussed in $\S$ 3. In general, the larger the BG subtraction, the steeper the index. The index derived without BG subtraction provides the lower limit.

\subsection{Fitting the Correlation}
We apply a pixel-to-pixel analysis and derive the power index of the K-S law. We use data above 3$\sigma$ significance in both $\Sigma_{\text{H$_2$}}$ and $\Sigma_{\text{SFR}}$. We adopt two sampling sizes, i.e, 750 and 500 pc; the former is the scale used in the previous CO($J$=2-1) study (B08), and the latter is a resolution similar to the one in \citet{ken07}. These pixel sizes are much larger than the point spread function (PSF) or beam size of 24 $\mu$m ($\sim$ 6$\arcsec$) and CO (1.9$\arcsec$ $\sim$ 4.8$\arcsec$) images.

We fit the data in a logarithmic space:
	\begin{equation}
		\log{(\Sigma_{\text{SFR}})} = A_{\text{fit}} + N \times \log{(\Sigma_{\text{H$_2$}})}    ,
	\end{equation}
where $A_{\text{fit}}$ is an intercept and $N$ is the index of the K-S law. We use the $FITEXY$ routine (\citealp{pre92}), which accounts for measurement errors along both $x$ and $y$ axes, thus provides more robust regression results, and has been used in previous studies (\citealp{ken07,rah11}). When comparing with B08, we also use the $OLS$ $bisector$ method (\citealp{iso90}) adopted by B08, which returns a bisector line in $x-$ and $y-$axes without the errors taken into account. We note that the choice of CO threshold surface density (i.e., 3$\sigma$) affects little; we repeated the analysis below with various thresholds and obtained the same conclusions.

\section{RESULTS}
\begin{figure}
\begin{center}
\includegraphics[width=18cm]{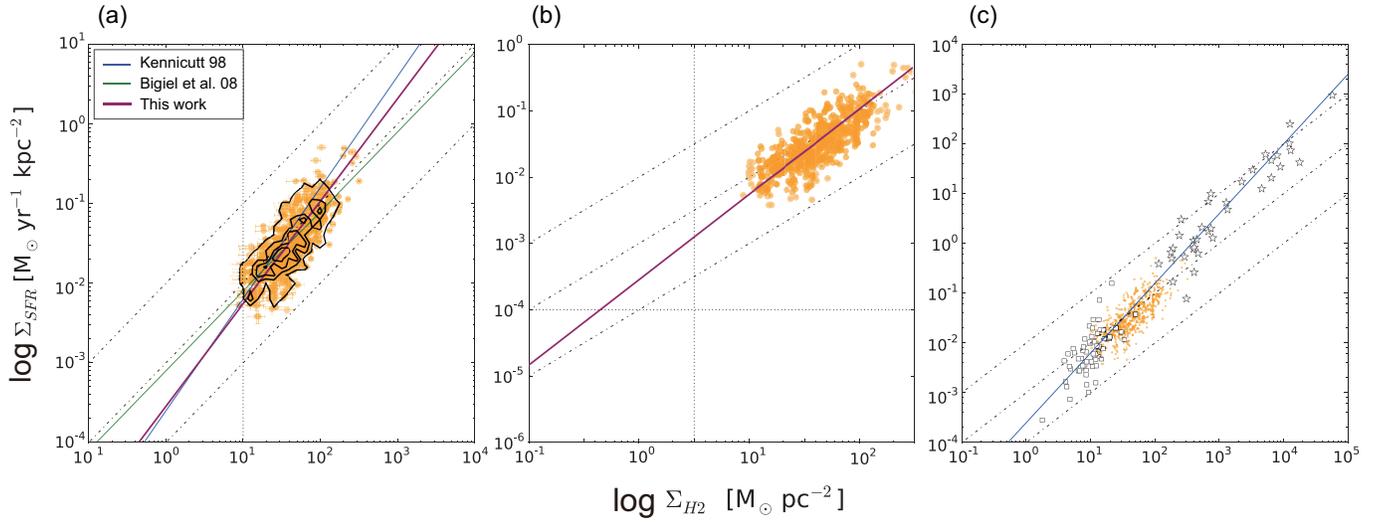}
\caption{K-S plots, without BG subtraction, on 750 pc scale. Data above 3$\sigma$ (yellow points) in our 10 sample galaxies are included (a), and on the same dynamical ranges of B08 (b) and K98a,b (c). White boxes and stars in (c) show the location of normal disk and nuclear starburst in K98a,b, respectively. Dot-dashed lines are SFE of 10$^{-8}$, 10$^{-9}$ and 10$^{-10}$ yr$^{-1}$ from top to bottom. Green and blue lines are the best fit of B08 and K98a,b, respectively. The light purple straight line is best fit through our data. Overlaid black contours are density distributions of the data points. Figures in color can be found in the on-line edition.}
\label{fig:ksken}
\end{center}
\end{figure}

We smooth the CO($J$=1-0) data to a 750 pc resolution (i.e., matching the resolution for all galaxies at different distances) and fit the K-S relation. This is the resolution used in the CO($J$=2-1) study by B08, and our results can be directly compared with the linear correlation $N$ $\sim$ 1 of B08 (more precisely, $N$ = 1.0$\pm$0.1 and $A_{\text{fit}}$ = --2.1$\pm$0.2). We combine all the data points from all galaxies to fit a line using measurement errors as weights. B08 did not subtract the extended BG emission, and therefore, in our first analysis, we do not make the subtraction from either H$\alpha$ or 24 $\mu$m images. A fit with either $FITEXY$ or the $OLS$ $bisector$ provides a super-linear slope of $N$ =  1.3$\pm$0.06 and $A_{\text{fit}}$ = --3.6$\pm$0.06 (Figure \ref{fig:ksken}a), steeper than the linear slope of B08 (see Table \ref{tab:fit}). Our power-law index is consistent with those of K98a,b within the error ($N$ = 1.4$\pm$0.15, Figure \ref{fig:ksken}c).

\begin{figure}
\begin{center}
\includegraphics[width=18cm]{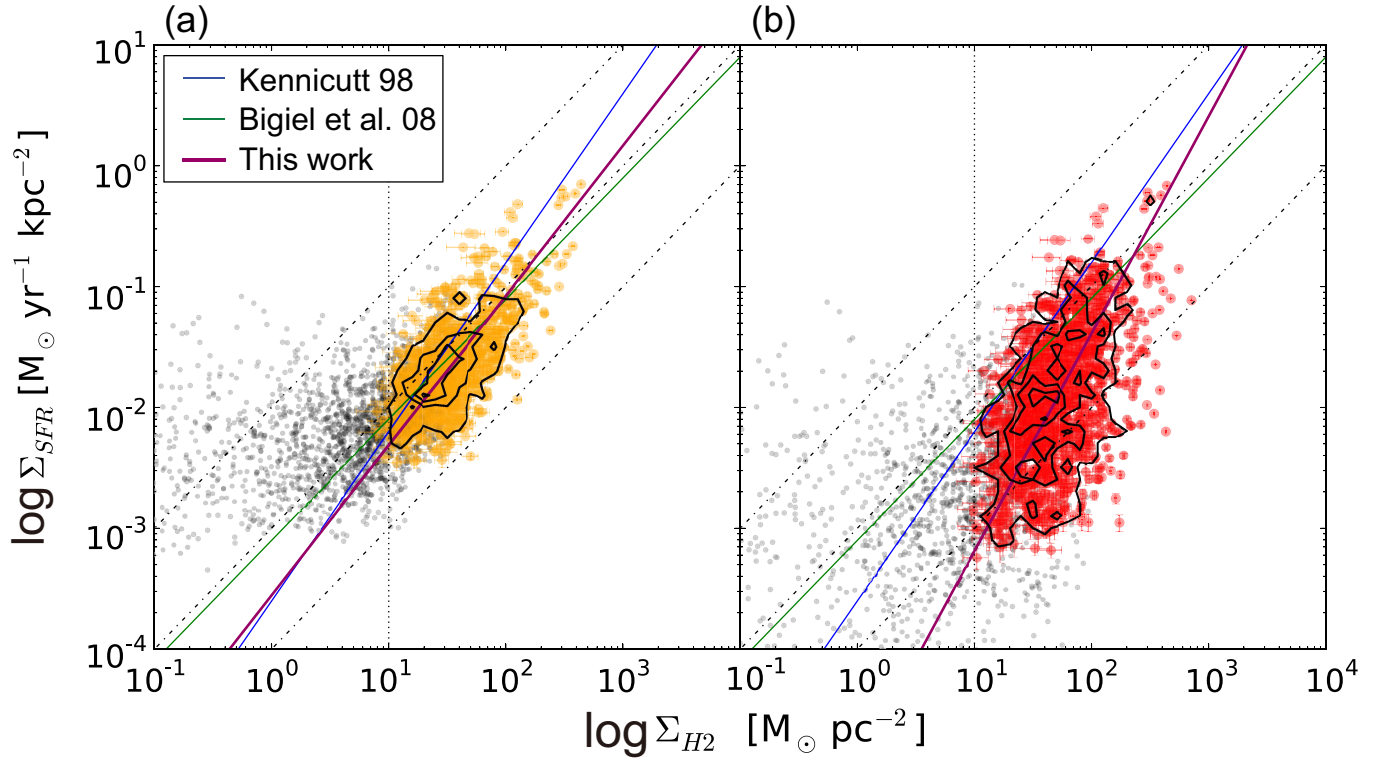}
\caption{K-S plot on 500 pc scale, (a) without and (b) with BG subtraction, for our 10 sample galaxies. Dot-dashed lines are SFEs of 10$^{-8}$, 10$^{-9}$ and 10$^{-10}$ yr$^{-1}$, from top to bottom. Green and blue lines are the best fit of B08 and K98a,b, respectively. Light purple lines are the best fit linear regressions through our data, yielding slopes of 1.3 (a) and 1.8 (b), respectively. Overlaid black contours are density distributions of the data points. We also plot all data points below 3$\sigma$ as black dots. Figures in color can be found in the on-line edition.}
\label{fig:ksall}
\end{center}
\end{figure}

The results stay the same even if a slightly higher resolution of 500 pc is used ($N$ = 1.3$\pm$0.06 and $A_{\text{fit}}$ = --3.6$\pm$0.09; Figure \ref{fig:ksall}a). If the BG emission is subtracted from H$\alpha$ and 24 $\mu$m (Figure \ref{fig:ksall}b), the slope becomes even steeper ($N$ = 1.8$\pm$0.10 and $A_{\text{fit}}$ = --5.0$\pm$0.17), more discrepant from the CO($J$=2-1) results. This trend -- a steeper slope after BG subtraction -- is consistent with that found by \citet{liu11} with a smaller sample (two galaxies). This is caused by systematic variations of the BG emission; the fraction of the emission becomes larger at lower $\Sigma_{\text{SFR}}$ regions. Some schemes of BG subtraction have been experimented (e.g. \citealp{rah11, liu11, ler12}), but these systematics -- greater reduction in flux in lower emission regions -- are general. The BG subtraction makes the K-S law slope steeper as more BG emission is subtracted. [Note that a potential presence of diffuse CO emission is suggested recently \citep{pet13}, though the survival and excitation mechanism of CO in the diffuse environment is still a subject on debate. As an experiment, we also imposed a local BG subtraction in the CO images as well as in H$\alpha$ and 24 $\mu$m and found no significant change (d$N$ $\sim$ 0.1).]

Our errors in $N$ and $A_{\text{fit}}$ are only statistical errors and do not include systematic errors. We adopted two fitting methods, FITEXY and OLS bisector, for comparisons with the previous work. \citet{bla09} discussed the disadvantage of these methods and adopted a Monte Carlo fitting method for a more realistic treatment of systematics and accurate determination of the parameters and errors \citep[see also ][]{she13}. Such an analysis is beyond the scope of this paper. We note, however, that in the end, our errors in $N$ ($\sim$ 0.06 and 0.10) are of the same order as those from the sophisticated analysis \citep[$\sim 0.05$;][]{bla09} while \citet{ler13} reported a slightly larger error $\sim 0.15$.

All our results with the CO($J$=1-0) data (at both 750 pc and 500 pc resolutions with and without BG subtractions) show the super-linear/power-law slope for the K-S law, inconsistent with the previous finding using CO($J$=2-1). 
CO($J$=1-0) data provide the super-linear correlation of $N$ $\sim$1.3--1.8.

\begin{table}
\caption{Fitted parameters of best fit linear regressions}
\begin{center}
\begin{tabular}{lcccc}
\tableline
	Scale		& $A_{\text{fit}}$	& $N$				\\
\tableline
	750 pc		& --3.6$\pm$0.06	& 1.3$\pm$0.06	\\
	500 pc (1)	& --3.6$\pm$0.09	& 1.3$\pm$0.06	\\
	500 pc (2) 	& --5.0$\pm$0.17	& 1.8$\pm$0.10	\\
\tableline
\end{tabular}
\end{center}
\footnote{500 pc scale (1) without and (2) with BG subtractions.} 
\label{tab:fit}
\end{table}

\section{DISCUSSION}
\subsection{The Cause of the Discrepancies}
We demonstrate that the CO($J$=1-0) data provide a super-linear slope of the K-S law, in contrast to the linear slope obtained by the CO($J$=2-1) studies, even when the diffuse BG emission is not subtracted from SFR tracer images. We here discuss possible causes of the discrepancy: differences in SFR tracers (i.e., H$\alpha$+24 $\mu$m vs FUV+24 $\mu$m), and differences in molecular gas density tracers (i.e., CO $J$ =1-0 vs 2-1). 

B08 estimated SFR using FUV and 24 $\mu$m, instead of H$\alpha$ and 24 $\mu$m. The two SFRs are compared and found consistent by previous studies. For example, some studies (B08; \citealp{ler08, liu11,ler12}) demonstrated that the $\Sigma_{\text{SFR}}$ from FUV+24 $\mu$m are equal to those derived from H$\alpha$+24 $\mu$m by \citet{cal07}, though they did not account for the BG emission. They also noted that the correlation becomes poorer in the range of very low SFR ($\Sigma_{\text{SFR}}$ $\leq$ 10$^{-3}$ M$_\odot$ yr$^{-1}$ kpc$^{-2}$). Our analysis does not suffer from this, since all the points in Figure \ref{fig:ksall} are above this boundary SFR. These comparisons in the previous studies give us a confidence that the difference in SFR tracers are unlikely the cause of the discrepancy between our results and those of B08.

The difference between CO($J$ =1-0) and CO($J$ =2-1) is likely a contributor to the discrepancy both empirically and due to physical considerations. The observed kinetic temperature of molecular gas is typically $\sim$ 10 K (\citealp{sco87}), which is above the level energy temperature of 5.5 K for the $J$ = 1, but below the temperature of 16.5 K for $J$ = 2. Therefore, a slight change in gas kinetic temperature affects the excitation for CO($J$=2-1) emission significantly. Furthermore, the change of the molecular gas volume density also affects the excitation for CO($J$=2-1). The different critical densities of the transitions ($\sim$ 10$^{3-4}$ cm$^{-3}$ and a few $\times$10$^{2}$ cm$^{-3}$ for $J$=2-1 and 1-0, respectively) make their ratio sensitive to local gas density. In fact, the ratio of CO($J$=2-1) to CO($J$=1-0) varies systematically with SFR and $\Sigma_{\text{gas}}$ in M 51 \citep{kod12, vla13} and in the Galaxy (e.g. \citealp{sak95, sak97, saw01}), although this systematic trend was often buried in noise \citep{ler13}. The higher SFR and $\Sigma_{\text{gas}}$ are, the greater the ratio is; this systematic trend directly affects the study of the K-S law using CO(2-1).

Previous studies suggested a change of the K-S law index when a higher CO rotational transition is used (e.g. \citealp{nar08, bay09}). Indeed, the excitation condition depends on the volume density and temperature; for example, the fraction of the gas above critical density of a line excitation is one of the key determinants for the slope of the K-S law (\citealp{nar08}). If the critical density for a line emission is high and close to the conditions of star-forming gas, that line emission can naturally show a linear relationship with SFR. Therefore, a higher-$J$ CO line transition tend to show a linear slope ($N$ = 1, e.g. \citealp{ion09}). On the other hand, CO($J$=1-0) has a relatively lower critical density for excitation, tracing the bulk molecular gas. Our results of super-linear correlation by CO($J$=1-0) implies that the SFE depends on the average gas surface density over $\sim$ 1 kpc scale, indicating the importance of environments around GMC on star formation.

\subsection{Implications for Star Formation}
The super-linear slope of the K-S law perhaps indicates that the process of star formation is nonlinear when it is seen on $\geq$ 500 pc scales. Our results with CO($J$=1-0) suggest a super-linear correlation both at 500 pc and 750 pc scales no matter whether the diffuse BG emission is subtracted or not from SFR tracer images. These spatial scales are much larger than the typical GMC size ($\sim$ 40 pc; \citealp{sco87}). The super-linear correlation at these scales may suggest some nonlinearity in processes of dense gas core formation and star formation in individual GMCs (e.g., \citealp{cal12}).

A relatively simple model of star formation can reproduce the super-linear slope $N$. The self-gravity of the gas naturally explains $N$ = 1.5 since the free-fall time $\tau_{\text{ff}}$ is proportional to the inverse of square root of the density. $\Sigma_{\text{SFR}}$ should be proportional to $\Sigma_{\text{gas}}$/$\tau_{\text{ff}}$ (e.g. \citealp{kru05, kru12}). Gravitational instability can play a role at larger scales. For example, a consideration of Toomre's Q instability parameter in rotating galactic disks also leads to $N$ = 1.5 (e.g. \citealp{sil97, elm02}). A steeper slope $N$ = 2 is possible as well if star formation is triggered by cloud-cloud collisions (e.g. \citealp{tan00, tan10}). Our result suggests that the star formation efficiency is not constant across galactic disks, though the non-linear process that plays a dominant role in driving star formation remains difficult to identify through our works.

\acknowledgments
We are grateful to the referee for constructive comments to improve this paper. We thank to Frank Bigiel, Rahul Shetty and Junichi Baba for discussions, Yasutaka Kurono for helping us to combine our CO($J$=1-0) data, and James Barrett for helpful comments on the English. We also thank  the SINGS team, the NRO staff for NRO45 observations, and the CARMA staff for CARMA observations. Support for CARMA construction was derived from the Gordon and Betty Moore Foundation, the Kenneth T. and Eileen L. Norris Foundation, the James S. McDonnell Foundation, the Associates of the California Institute of Technology, the University of Chicago, the states of California, Illinois, and Maryland, and the National Science Foundation. Ongoing CARMA development and operations are supported by the National Science Foundation under a cooperative agreement and by the CARMA partner universities. This research was partially supported by Hayakawa Yukio Foundation. JK acknowledges support from the NSF through grant AST-1211680.


\begin{thebibliography}{}
\bibitem[Bigiel et al. (2008)]{big08} Bigiel, F. et al. 2008, \aj, 136, 2846 (B08)
\bibitem[Blanc et al. (2009)]{bla09} Blanc, G. A., Heiderman, A., Gebhardt, K., Evans, II, N. J. $\&$ Adams, J., 2009, \apj, 704, 842
\bibitem[Bayet et al. (2009)]{bay09} Bayet, E., Gerin, M., Phillips, T. G., $\&$ Contursi, A., 2009, \mnras, 399, 264
\bibitem[Bolatto et al. (2008)]{bol08} Bolatto, A. D., Leroy, A. K., Rosolowsky, E., Walter, F., $\&$ Blitz, L., 2008, \apj, 686, 948
\bibitem[Bolatto, Walfire \& Leroy (2013)]{bol13} {Bolatto}, A.~D. and {Wolfire}, M. and {Leroy}, A.~K. \ 2013, astro-ph/1301.3498
\bibitem[Calzetti et al. (2007)]{cal07} Calzetti, D. et al. 2007, \apj, 666, 870
\bibitem[Calzetti, Liu $\&$ Koda (2012)]{cal12} Calzetti, D., Liu, G., $\&$ Koda, J., 2012, \apj, 752, 98
\bibitem[Dale et al. (2009)]{dal09} Dale, D. A. et al. 2009, \apj, 703, 517
\bibitem[Dame, Hartmann $\&$ Thaddeus (2001)]{dam01}Dame, T. M., Hartmann, D., $\&$ Thaddeus, P., 2001, \apj, 547, 792
\bibitem[Donovan Meyer et al. (2012)]{don12} Donovan Meyer, J., et al. 2012, \apj, 744, 42
\bibitem[Donovan Meyer et al. (2013)]{don13} Donovan Meyer, J., et al. 2013, ApJ accepted (astro-ph1305.5275).
\bibitem[Draine $\&$ Li (2007)]{dra07} Draine, B. T. $\&$ Li, A., 2007, \apj, 657, 810 
\bibitem[Draine et al. (2007)]{dra07a} Draine, B. T., et al. 2007, \apj, 663, 866
\bibitem[Elmegreen (2002)]{elm02} Elmegreen, B. G., 2002, \apj, 577, 206
\bibitem[Ferguson et al. (1996)]{ferg96} Ferguson, A. M. N., Wyse, R. F. G., Gallagher, J. S., III $\&$ Hunter, D. A., 1996, \aj, 111, 2265
\bibitem[Iono et al. (2009)]{ion09} Iono, D. et al. 2009, \apj, 695, 1537
\bibitem[Isobe et al. (1990)]{iso90} Isobe, T., Feigelson, E. D., Akritas, M. G., $\&$ Babu, G. J., 1990, \apj, 364, 104
\bibitem[Kennicutt (1998a)]{ken98a} Kennicutt, Jr., R. C., 1998, \araa, 36, 189 (K98a)
\bibitem[Kennicutt (1998b)]{ken98b} Kennicutt, Jr., R. C., 1998, \apj, 498, 541 (K98b)
\bibitem[Kennicutt et al. (2003)]{ken03} Kennicutt, Jr., R. C. et al. 2003, \pasp, 115, 928
\bibitem[Kennicutt et al. (2007)]{ken07} Kennicutt, Jr., R. C. et al. 2007, \apj, 671, 333
\bibitem[Kennicutt $\&$ Evans (2012)]{ken12} Kennicutt, Jr., R. C., $\&$ Evans, N. J. II., 2012, \araa, 50, 531
\bibitem[Knapen et al. (2004)]{kna04} Knapen, J. H. et al. 2004, \aap, 426, 1135
\bibitem[Koda et al. (2011)]{kod11} Koda, J. et al. 2011, \apjs, 193, 19
\bibitem[Koda et al. (2012)]{kod12} Koda, J. et al. 2012, \apj, 761, 41
\bibitem[Krumholz $\&$ McKee (2005)]{kru05} Krumholz, M. R. $\&$ McKee, C. F., 2005, \apj, 630, 250
\bibitem[Krumholz, Dekel $\&$ McKee (2012)]{kru12} Krumholz, M. R., Dekel, A. $\&$ McKee, C. F., 2012, \apj, 745, 69
\bibitem[Leroy et al. (2008)]{ler08} Leroy, A. K. et al. 2008, \aj, 136, 2782
\bibitem[Leroy et al. (2012)]{ler12} Leroy, A. K. et al. 2012, \aj, 144, 3
\bibitem[Leroy et al. (2013)]{ler13} {Leroy}, A. K. et al. 2013, astro-ph/1301.2328.
\bibitem[Liu et al. (2011)]{liu11} Liu, G., Koda, J., Calzetti, D., Fukuhara, M. $\&$ Momose, R., 2011, \apj, 735, 63
\bibitem[Moustakas et al. (2010)]{mou10} Moustakas, J., Kennicutt, Jr., R. C., Tremonti, C. A., Dale, D. A., Smith, J-D. T., $\&$ Calzetti, D., 2010, \apjs, 190, 233
\bibitem[Narayanan et al. (2008)]{nar08} Narayanan, D., Cox, T. J., Shirley, Y., Dave, R., Hernquist, L. $\&$ Walker, C. K., 2008, \apj, 684, 996
\bibitem[Pety et al.(2013)]{pet13} {Pety}, J. and {Schinnerer}, E. and {Leroy}, A.~K. et al. \ 2013, astro-ph/1304.1396.
\bibitem[Press et al. (1992)]{pre92} Press, W. H., Teukolsky, S. A., Vetterling, W. T. $\&$ Flannery, B.P., 1992, Numerical recipes in FORTRAN. The art of scientific computing
\bibitem[Rahman et al. (2011)]{rah11} Rahman, N. et al. 2011, \apj, 730, 72
\bibitem[Rahman et al. (2012)]{rah12} Rahman, N. et al. 2012, \apj, 745, 183
\bibitem[Sakamoto et al. (1995)]{sak95} Sakamoto, S., Hasegawa, T., Hayashi, M., Handa, T., $\&$ Oka, T., 1995, \apjs, 100, 125
\bibitem[Sakamoto et al. (1997)]{sak97} Sakamoto, S., Hasegawa, T., Handa, T., Hayashi, M., $\&$ Oka, T., 1997, \apj, 486, 2765
\bibitem[Sawada et al. (2001)]{saw01} Sawada, T. et al. 2001, \apjs, 136, 189
\bibitem[Schmidt (1959)]{sch59} Schmidt, M., 1959, \apj, 129, 243
\bibitem[Schruba et al. (2011)]{sch11} Schruba, A. et al. 2011, \aj, 142, 37
\bibitem[Scoville $\&$ Sanders (1987)]{sco87} Scoville, N. Z. $\&$ Sanders, D. B., 1987, ASSL, 134, 21
\bibitem[Scoville et al. (2001)]{sco01} Scoville, N. Z., Polletta, M., Ewald, S., Stolovy, S. R., Thompson, R., $\&$ Rieke, M., 2001, \aj, 122, 3017
\bibitem[Silk (1997)]{sil97} Silk, J., 1997, \apj, 481, 703
\bibitem[Shetty, Kelly $\&$ Bigiel (2013)]{she13} Shetty, B., Kelly, B. C., $\&$ Bigiel, F., 2013, \mnras, 430, 288
\bibitem[Strong $\&$ Mattox (1996)]{sto96} Strong, A. W., $\&$ Mattox, J. R., 1996, \aap, 308, L21
\bibitem[Tan (2000)]{tan00} Tan, J. C., 2000, \apj, 536, 173
\bibitem[Tan (2010)]{tan10} Tan, J. C., 2010, \apj, 710, L88
\bibitem[Thilker, Braun $\&$ Walterbos (2000)]{thi00} Thilker, D. A., Braun, R., $\&$ Walterbos, R. A. M., 2000, \aj, 120, 3070
\bibitem[Verley et al. (2009)]{ver09} Verley, S., Corbelli, E., Giovanardi, C. $\&$ Hunt, L. K., 2009, \aap, 493, 453
\bibitem[Vlahakis et al.(2013)]{vla13} {Vlahakis}, C. and {van der Werf}, P. and {Israel}, F.~P. and {Tilanus}, R.~P.~J. \ 2013, astro-ph/1304.7408
\bibitem[Wong $\&$ Blitz (2002)]{won02} Wong, T., $\&$ Blitz, L., 2002, \apj, 569, 157
\end{thebibliography}
\end{document}